\documentclass[a4paper, 11pt]{article}
\usepackage[sort&compress,square,comma,authoryear]{natbib}
\usepackage{amsmath,amssymb}
\usepackage{graphicx}

\begin{document}

\begin{center}
{\bf \large First observations of irregular surface of interplanetary shocks at ion scales by Cluster}
\vspace{1cm}\\






Primo\v{z} Kajdi\v{c}(1), Luis Preisser(1), X\'ochitl Blanco-Cano(1), David Burgess(2), Domenico Trotta(2)\\
(1) Instituto de Geof\' isica, Universidad Nacional Aut\'onoma de M\'exico, Circuito de la investigaci\'on Cient\' ifica s/n, Ciudad Universitaria, Delegaci\'on Coyoac\'an, C.P. 04510, Mexico City, Mexico\\
(2) School of Physics and Astronomy, Queen Mary University of London, London E1 4NS, UK\\
\end{center}

We present the first observational evidence of the irregular surface of interplanetary (IP) shocks by using multi-spacecraft observations of the Cluster mission. In total we discuss observations of four IP shocks that exhibit moderate Alfv\'enic Mach numbers (M$_A\leq$6.5). Three of them are high-$\beta$ shocks with upstream $\beta$ = 2.2--3.7. During the times when these shocks were observed, the Cluster spacecraft formed constellations with inter-spacecraft separations ranging from less than one upstream ion inertial length (d$_i$) up to 100~d$_i$. Expressed in kilometers, the distances ranged between 38~km and $\sim$10$^4$~km. We show that magnetic field profiles and the local shock normals of observed shocks are very similar when the spacecraft are of the order of one d$_i$ apart, but are strikingly different when the distances increase to ten or more d$_i$. We interpret these differences to be due to the irregular surface of IP shocks and discuss possible causes for such irregularity.
We strengthen our interpretation by comparing observed shock profiles with profiles of simulated shocks. The latter had similar characteristics (M$_A$, $\theta_{BN}$, upstream ion $\beta$) as observed shocks and the profiles were obtained at separations across the simulation domain equivalent to the Cluster inter-spacecraft distances.


\section{Introduction}
Interplanetary (IP) shocks are ubiquitous in the heliosphere and are mostly associated with interplanetary coronal mass ejections \citep[ICME, e.g. ][]{sheeley:1985} and stream interaction regions \citep[SIR, e.g. ][]{gosling:1999}.
Since the solar wind (SW) is a collisionless plasma, the heliospheric shocks are also collisionless. \citet{marshall:1955} realized that purely resistive dissipation mechanisms cannot sustain collisionless shocks when their Mach number exceeds some critical value M$_{c}$. \citet{sagdeev:1966} suggested that particle acceleration and reflection at the shock surface could be an efficient mechanism to dissipate the kinetic energy of the incoming SW. 

Numerical simulations show that one of the characteristics of supercritical collisionless shocks is the nonstationarity or reformation of their surface which leads to its irregularity. Depending on the shock properties (Alfv\'enic Mach number M$_A$, the angle $\theta_{BN}$ between the upstream B-field and the shock normal and the ratio $\beta$ of the upstream thermal to magnetic pressure), the nonstationarity may be due to the self-reformation of the shock surface \citep[e.g., ][]{biskamp:1972, leroy:1982}, the whistler mode waves emitted in the foot/ramp \citep[][]{scholer:2007, hellinger:2007, lembege:2009} or the upstream ultra-low frequency (ULF) waves and steepened foreshock structures \citep[e.g., ][]{burgess:1989b, schwartz:1991, kraussvarban:1991,kraussvarban:2008}.

Self-reformation is favoured by large M$_A$ and small upstream $\beta$ and produces shock rippling at small spatial scales along the shock surface ($\lesssim$1 ion inertial length, d$_i$). Large amplitude whistler waves emitted in the foot cause rippling at spatial scales of a few d$_i$. Finally, irregularities due to upstream ULF waves occur at spatial scales of several tens of d$_i$.

Shock surface rippling was first studied by comparing with 2D hybrid (kinetic ions, masseless fluid electrons) simulations by \citet{winske:1988} and later by, for example, \citet{lowe:2003} and \citet{ofman:2013}. The latter showed that the difference between the global shock normal (determined by the far upstream and far downstream states) and the local normal (determined by the local direction of the fastest variation of the magnetic field) due to rippling can be as large as 40$^\circ$. \citet{burgess:2006a,burgess:2006b} studied simulated high Mach number (M$_A$=9.7), almost perpendicular ($\theta_{BN}\lesssim$90$^\circ$) shocks and showed that the B-field profiles observed by virtual probes varied even when the probes were closely separated ($\leq$2.5~d$_i$) and that the shock surface ripples cause large variations of the local $\theta_{BN}$.

Larger-scale irregularity of shock surface was reproduced by \citet{kraussvarban:2008} who performed local hybrid simulations of a shock with M$_A$=4.7 and $\theta_{BN}$=50$^\circ$. They observed that in the case of large-scale shocks, such as IP shocks, even a very small amount of reflected ions generate upstream compressional waves that bend the B-field lines and change the local $\theta_{BN}$. The portions of the shock that become more parallel eject more protons back upstream and further enhance the compressional waves there. These regions were found to travel along the shock surface leading to irregularities with a wavelength of 100~d$_i$ or more. 

Shock ripples and their importance for particle acceleration and for formation of downstream phenomena have been the subject of several works \citep[i.e., ][]{gedalin:2001, yang:2011, yang:2018, hietala:2009,hao:2016}. Rippling was observed at the Earth's bow-shock by several authors by using multi-spacecraft data from Cluster \citep{horbury:2001, horbury:2002, moullard:2006, lobzin:2008} and Magnetospheric Multiscale (MMS) mission \citep{johlander:2016,gingell:2017}. As for the IP shocks, several authors \citep[i.e., ][]{russell:1984, russell:2000, szabo:2001, szabo:2003, szabo:2005, pulupa:2008, koval:2010, kajdic:2017a} discussed and/or observed their non-planar structure at scales of several tens of Earth radii (R$_E$), but never at ion scales.

Although the particle acceleration mechanisms at Earth's bow-shock and at IP shocks should be the same, the IP shocks at 1~au have much larger curvature radii ($\sim$0.5~a.u.) and the majority exhibit smaller magnetosonic Mach numbers \citep[up to four, e.g. ][]{blancocano:2016}.  This is reflected in the way the ions are accelerated at IP shocks. For example, \citet{kajdic:2017a} reported first observations of field-aligned ions upstream of an IP shock with much higher energies that those observed at the Earth's bow-shock \citep[e.g., ][]{gosling:1978}.

Here we use the multi-spacecraft capabilities of the Cluster mission to study the structure of IP shock fronts. At the times of the shocks the four probes were separated between several tens to several thousands of kilometers, which corresponds to less than one to several tens of upstream ion inertial lengths (d$_i$). All the shocks exhibit moderate Alfv\'enic Mach numbers (M$_A\leq$6.5). Three of the shocks also exhibit upstream ion $\beta$ values (ratio between ion thermal and magnetic pressures) larger than unity. This is different from the studies of the rippling of the Earth's bow-shock surface where $\beta$ was either larger than 1 but M$_A$ were also very large \citep[e.g., ][]{moullard:2006} or $\beta$ was much smaller than 1 and M$_A$ was moderate \citep[e.g., ][]{johlander:2016, yang:2018}.

We compare shock profiles, local shock normals and geometries at each spacecraft and show that these vary even at small ($<$10~$d_i$) spacecraft separations. We attribute these differences to irregular IP shock surface and further strengthen our case with 2D hybrid simulations.

\section{Observations: Shock profiles}
In this section we compare B-field  profiles of four shocks observed on 17 January 2001, 5 April 2010, 26 February 2012 and 8 March 2012. We use data obtained by the Flux Gate Magnetometers \citep[FGM, ][]{balogh:1997, balogh:2001} onboard the four identical Cluster \citep{escoubet:1997} probes.

Figure~\ref{fig:shocks} shows Cluster FGM data with time resolution of 22 vectors per second at four different times during which the shocks were observed. Different colors correspond to different spacecraft with black, blue, green and red lines representing the Cluster~1 (C1), 2 (C2), 3 (C3) and 4 (C4) data, respectively. The IP shocks were selected from the {\it Catalog of IP shocks observed in the Earth's neighborhood by multiple spacecraft} available at http://usuarios.geofisica.unam.mx/primoz/IPShocks.html. Basic shock parameters are exhibited on each panel: the range of angles $\theta_{BN}$, the shock's Alfv\'enic (M$_A$) and magnetosonic (M$_{ms}$) Mach numbers (if all the data are available, otherwise only M$_A$ is provided) and the upstream ion $\beta$. If the data were missing in the catalog, values were obtained from the {\it Comprehensive database of interplanetary shocks} at www.ipshocks.fi.

During these times the Cluster probes formed different constellations. Separations of pairs of spacecraft are provided in Table~\ref{tab:mva}.

\subsection{17 January 2001 shock}
On 17 January 2001 a fast forward shock (panel a in Figure~\ref{fig:shocks}) was detected at 16:27:48~UT by the C1 and C3 spacecraft. C2 detected the shock about a second earlier, while C4 observed the initial B-field increase at the same time as C1 and C3, but the increase was much more gradual in its data. We see from Table~\ref{tab:mva} that in this case the spacecraft separations were between 530~km (7.4~d$_i$) and 1300~km (18.1~d$_i$).

C2 was the most sunward spacecraft, therefore detecting the shock first. It was followed by C3 and C4 while C1 was closest to the Earth. Although C3 and C4 had the most similar X$_{GSE}$ coordinates it was C1 and C3 that detected the shock ramp simultaneously. The spacecraft positions and the local shock orientation in the Y$_{GSE}$-Z$_{GSE}$ plane probably account for this.

We find that shock profiles observed by the four spacecraft are not the same (Figure~\ref{fig:shocks}a). C2 (blue) observed a ramp, an overshoot and possibly a foot. A small peak just upstream of the shock ramp could be a small whistler wave precursor. The B-field magnitude of the overshoot was 7.2~nT, which is more than in the other three cases. C1 and C3 (black and green) observed the shock ramp simultaneously but the B-field magnitudes were slightly different (7.0~nT and 6.5~nT, respectively). Just after the ramp, both spacecraft observed an overshoot, a short lasting undershoot and another peak with similar B-field magnitude value as the first peak. Finally, the shock ramp observed by C4 (red) is not as steep as in the other three cases. After the overshoot there is a strong dip (undershoot) after which the B-field magnitude rises again. Just before the shock ramp, C1 (black) observes a small peak similar to that observed by C2.
C4 does not observe any foot, however it detects a distinct compressive structure which lasts $\sim$2~s peaking at 16:27:46~UT. Similar structures are observed by C1 at 16:27:35.5~UT and 16:27:42.5~UT.

\subsection{5 April 2010 shock}
This shock was detected at 08:24:59~UT by C1 (Figure~\ref{fig:shocks}b). C2 detected it about two seconds earlier while C3 and C4 observed it simultaneously some 4 seconds later. The positions of the spacecraft were such that C3 and C4 were separated by only 200~km (1.5~d$_i$), so they observed almost identical shock profiles. These two probes observe a well defined ramp and overshoot followed by two dips and peaks. The latter could be old overshoots from previous reformation cycles or compressive waves. C3 detected a short lasting ($\sim$1~s) whistler precursor with frequency of $\sim$3~Hz which can barely be distinguished in C4 data. C2 (blue) observed a steep ramp and an overshoot followed by a very short lasting dip and another increase of B. Upstream of the ramp C2 observed short lasting whistler precursor. Further downstream there is a deep dip in the C2 data, similar to that observed by C3 and C4. C1 (black) observed a much more gradual shock transition, typical of quasi-parallel ($\theta_{BN}\leq$45$^\circ$) shocks.

\subsection{26 February 2012 shock}
C1 observed this shock at 21:38:18~UT (Figure~\ref{fig:shocks}c) followed by C3 and C4 spacecraft that observed it simultaneously $\sim$3~s later while C2 observed it $\sim$5~s later. C3 and C4 were separated by only 38~km (0.37~d$_i$) while the other separations were much larger. It is interesting that while all $\theta_{BN}$ values (13$^\circ$-39$^\circ$) indicate that this is a quasi-parallel shock, all profiles resemble those of quasi-perpendicular ($\theta_{BN}>$45$^\circ$) shocks with steep ramps and overshoots. C1 and C2 observed a high frequency whistler precursor just upstream of the shock. These waves exhibit small amplitudes (0.1~nT - 0.2~nT) and frequencies of $\sim$2~Hz. In the case of C1 the precursor lasts for $\sim$1.5~s while in C2 data it lasts for $\sim$7~s. C2, C3 and C4 profiles show three well defined peaks separated by two dips. Although these peaks could be some compressive waves they could also be overshoots from previous reformation cycles (see Section~\ref{sec:shocknorm}). In the case of C2 the first peak is actually separated from the shock ramp and exhibits higher B-field magnitude than the actual overshoot. C1 profile also shows these features but with much smaller amplitudes.

\subsection{8 March 2012 shock}
Our last shock was observed by C1 at $\sim$11:02:43~UT (Figure~\ref{fig:shocks}d) then by C2 about 3 seconds later and lastly by C3 and C4 spacecraft simultaneously. C3 and C4 were separated by 55~km (0.92~d$_i$) while the other separations were much larger (Table~\ref{tab:mva}). C3 and C4 profiles are thus almost identical.  All spacecraft observe whistler mode precursors upstream of the shock ramp but their appearance varies from spacecraft to spacecraft. In C1 data the whistlers lasted for $\sim$7~s featuring at least three wave-trains, while the other three spacecraft observed one wave-train during $\sim$5~s time interval. Frequencies of these waves were $\sim$2~Hz and their amplitudes increased with proximity to the shock ramp. The shock ramp was steepest in C1 data, while the other profiles show whistler waves inside the shock ramps.

\section{Observations: Local shock normals}
\label{sec:shocknorm}
The $\theta_{BN}$ angles shown in Figure~\ref{fig:shocks} were obtained using the magnetic coplanarity theorem. This requires averaging of upstream and downstream fields during chosen time intervals (but exclude the shock transition) which are then used to calculate the shock normal and $\theta_{BN}$. Thus one obtains some time-averaged values. When multiple inter-spacecraft separations are small ($\lesssim$100~d$_i$), one would expect the shock normals calculated this way to coincide within the margin of error. This is because nonstationarity is quasi-cyclic so local shock normals and $\theta_{BN}$ vary in time around some average value (see Section~\ref{sec:simulations}). In order to study shock surface irregularity, we need local shock normals at the times when the shocks were observed by each spacecraft and see how they vary as a function of inter-spacecraft separation.

For this we use a novel one-spacecraft method based on the shock normal coordinate system (SNCS). The latter contains three perpendicular axes, $n$, $l$ and $m$. The $n$-axis is parallel to the shock normal, the $l$-axis contains a projection of the upstream B-field on the shock plane, while the $m$-axis completes the right-hand coordinate system. In this coordinate system only B$_l$ component changes from upstream to downstream.

This is of course strictly  true only for MHD shocks. In the case of collisionless shocks there exist an out-of-plane component of the magnetic field produced in the shocks's foot and overshoot. However the largest variation of the B-field is still produced due to the shock ramp itself and it occurs in the $l$ direction.

In order to find the SNCS using given interval, we first smooth the B-field data by using a 4-second sliding window in order to remove the upstream whistlers. We then perform minimum variance analysis \citep[MVA, ][]{sonnerup:1998} of the B-field across the shock and postulate that the direction of maximum variance gives us the $l$-direction. We also obtain two more vectors, perpendicular to $l$. We then rotate one of them around the $l$-axis and calculate the absolute value of the mean of the B-field projection along it. Once this value reaches its minimum close to 0, we take the corresponding vector to point along the $m$-axis and the remaining vector has to point along $n$. 

This method is not without errors. Their main sources are the intrinsic error of the MVA and the choice of time intervals used for the MVA. Details on how the errors were estimated and examples of B-field profiles of shocks in the SNCS are provided in the Appendix A and in the supplement repository available at https://zenodo.org/record/2587992.

Figure~\ref{fig:scatter} shows angles between pairs of normals ($\theta_{NN}$) as a function of spacecraft separation in units of upstream d$_i$. The results are summarized in Table~\ref{tab:mva}. It can be seen that $\theta_{NN}$ angles match very well at small spacecraft separations. As long as the spacecraft are $\lesssim$5~d$_i$ apart, their normals are within $\leq$4$^\circ$. As the spacecraft separation increases to $\sim$100~d$_i$, the $\theta_{NN}$ angles grow to $\sim$30$^\circ$.

Table~\ref{tab:mva} also shows local $\theta_{BN}$ angles. We estimate the errors of $\theta_{BN}$ to be $\sim\pm$5$^\circ$. These stem mostly from errors of normal directions and the errors due to selection of time intervals used for calculating the upstream B-field. It can be seen that the average $\theta_{BN}$ angles vary substantially from probe to probe.

\section{Simulations results}
\label{sec:simulations}
In order to further strengthen our case in favor of IP shock irregularity at ion scales, we use the 2D HYPSI numerical code \citep{burgess:2015, gingell:2017} to carry out two simulations of collisionless shocks.

We use a grid of N$_x\times$N$_y$ = 1000$\times$800 cell with cell size of 0.5~d$_i$ in both directions. The SW is injected at the left boundary with velocity of 3.0~V$_A$ and 4.5~V$_A$ resulting in M$_A\sim$4.4 and 6, respectively, for the shock. The upstream $\beta$ values were 2.4 and 0.2, respectively, which is close to the observed values. Density and B-field are normalized to their upstream values. Initially there were 100 simulation particles per cell. At the right boundary the particles are reflected while periodic boundary conditions are applied in the $y$ direction. The upstream B-field lies in the simulation plane at an angle of 50$^\circ$ with respect to the $x$-axis. This is also the value of the average $\theta_{BN}$ which we denominate as $\theta_{BN,0}$.

In Figure~\ref{fig:simulation}a we show simulation results for high-$\beta$, low M$_A$ shock at time t=222.5~$\Omega_i^{-1}$ ($\Omega_i$ is the upstream proton gyrofrequency) with x = [-25,25]~d$_i$ and y = [40, 120]~d$_i$. Results for high-M$_A$ shock can be seen in the Appendix B and in the supplement repository. $x$=0 is the average shock position obtained from the averaged (in $y$) B-field profile. The colors represent the B-field magnitude. The white curve marks the shock front determined by B$\geq$2.2 and then smoothed with eleven-cell (5.5~d$_i$) wide runing window. Finally, we calculate the local shock normals (blue arrows).

In the upstream region, there are compressive structures shaded in darker-orange shades. These features are convected towards the shock surface and are the primary cause of its reformation. To show this, four animations are available in the supplement repository. 
Two of those animations correspond to the Figure~\ref{fig:simulation}a). The long animation begins at t=102.5~$\Omega^{-1}$, ends at t=327.5~$\Omega^{-1}$ and is 30~s long. The shorter animation is 5~s long and shows the shock evolution durint t=220~~$\Omega^{-1}$ and t=232.5~~$\Omega^{-1}$. 
Another two animations correspond to the Figure~\ref{fig:simulation2}a), have durations of 5 and 36~s and show shock evolution during time intervals of 210-232~$\Omega^{-1}$ and 52.5-317.5~$\Omega^{-1}$, respectively.
In both cases, we can observe compressive upstream magnetic structures being convected towards the shock front. When they get really close and start touching the shock, their amplitudes rise sharply. They merge with the shock front and their upstream edges become new shock ramps.

Figure~\ref{fig:simulation}b shows four B-field profiles for $y$=50~d$_i$ (black), 60~d$_i$ (red), 65~d$_i$ (cyan) and 100~d$_i$ (magenta). The black profile exhibits a gradual rise and several downstream peaks and dips. The red profile shows upstream whistlers and a steep ramp. The cyan profile does not show any whistlers, only a steep ramp. The magenta profile exhibits a more gradual rise and a strong downstream dip. 

We can see in Figure~\ref{fig:simulation}a that the upstream variations seen in black, red and cyan profiles are due to upstream structures that have arrived close to the shock and/or the whistler precursors. The multiple dowstream peaks may be due to old overshoots from previous reformation cycles (red profile). The large dips seen in black and magenta profiles coincide with downstream regions with B values similar to those in the upstream region. These regions, although downstream of the shock, have not yet been fully compressed.

Figure~\ref{fig:simulation}c shows the distribution of $\theta_{BN}$ angles at time t=222.5~$\Omega_i$. The average and median ($\mu$) $\theta_{BN}$ angles are close to $\theta_{BN,0}$, but the local $\theta_{BN}$s have values anywhere between $\sim$10$^\circ$ and $\sim$90$^\circ$. 

Figure~\ref{fig:simulation}d shows the time evolution of the $\theta_{BN}$ for the point on the shock surface at y=100~d$_i$. We see that $\theta_{BN}$ oscillates around $\theta_{BN,0}$. The Fourier spectrum of the $\theta_{BN}$ variation (Figure~\ref{fig:simulation}e) reveals the presence of several periods.

Figure~\ref{fig:simulation}f shows $\theta_{NN}$ angles for all pairs of normals on panel a) as a function of distance. These angles increase and decrease due to the shock irregularities as seen on panel a. There is no real tendency between $\theta_{NN}$ and the distance, although $\theta_{NN}$ tend to be smaller at small distances.

The high-$M_A$, low-$\beta$ run (see Appendix B and the supplement repository) differs from the one discussed here mainly in that upstream and downstream compressive structures and the shock exhibit larger amplitudes, the standard deviation of the $\theta_{BN}$ distribution is larger, the periods in the $\theta_{BN}$ spectra are shorter, and the shock is more structured especially at smaller separations ($\lesssim$10~d$_i$), so there is even less correlation between $\theta_{NN}$ and the distance.

\section{Discussion and conclusions}
\begin{figure}[h]
 \centering
 \includegraphics[width=0.8\textwidth]{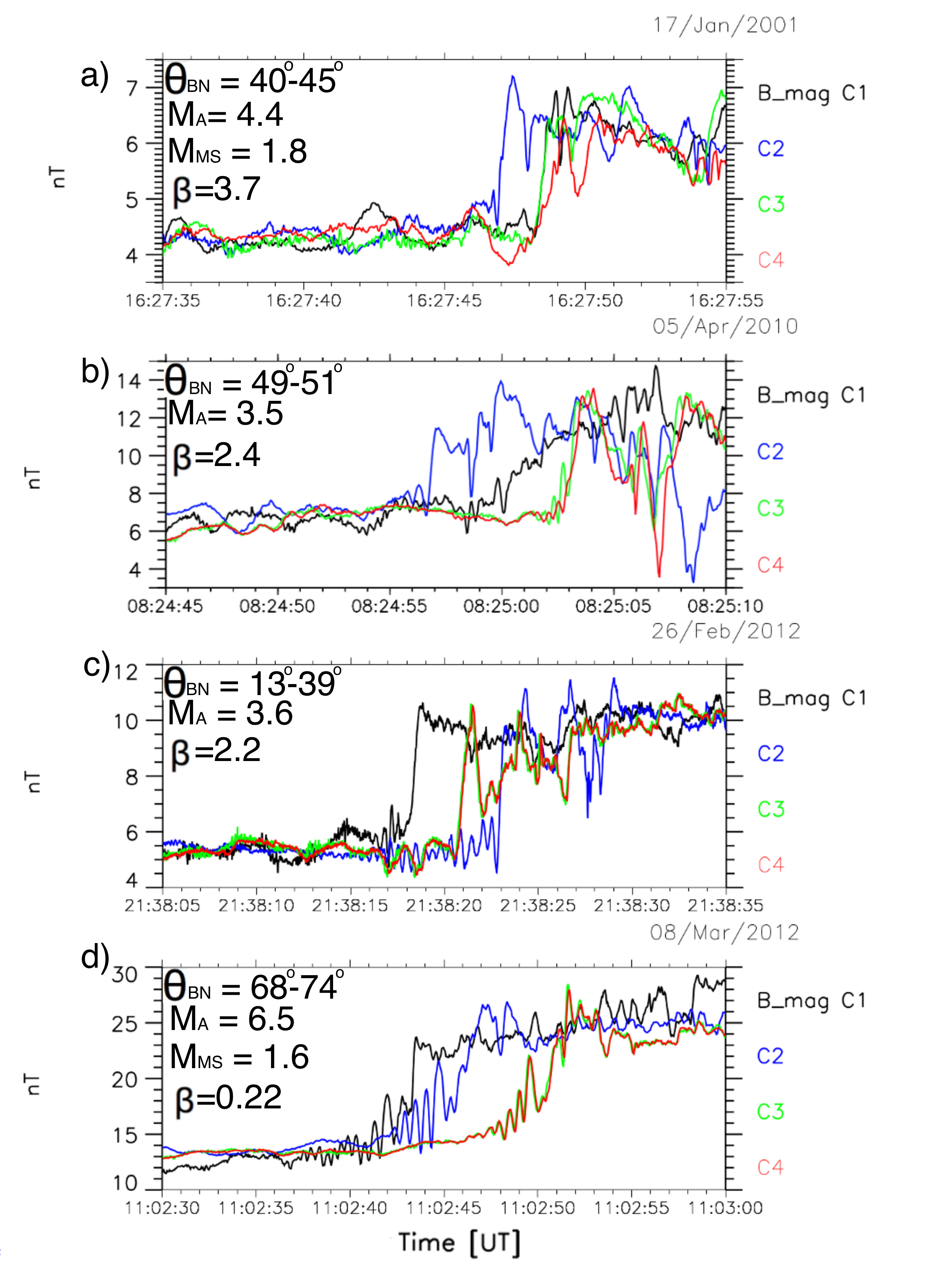}
 \caption{B-field profiles during four intervals when IP shocks were detected. The black, blue, red and green lines represent the data from C1, C2, C3 and C4 spacecraft, respectively.}
 \label{fig:shocks}
 \end{figure}

In this work we present the first direct observational evidence for {\bf an irregular} surface of IP shocks at ion scales. We show four case studies (Figure~\ref{fig:shocks}) that were observed by the four Cluster probes with inter-spacecraft separations between 38~km and $\sim$10$^4$~km (0.37~d$_i$-92~d$_i$). We show that B-field shock profiles vary from probe to probe. When the spacecraft were $\lesssim$1~d$_i$ apart (Table~\ref{tab:mva}), the profiles are very similar. 
When spacecraft are separated by more than 10~d$_i$ the shock profiles differ significantly. We attribute these differences to be associated with an {\bf irregular} shock front.

We further strengthen our argument by calculating local shock normals at each spacecraft for which we design a new one-spacecraft analysis method (see Section~\ref{sec:shocknorm}). We plot the angle between pairs of single spacecraft normals, $\theta_{NN}$, as a function of the distance between the probes. On average these angles tend to be smaller when the spacecraft are $\lesssim$5~d$_i$ apart and then increase as the distance increases up to $\sim$100~d$_i$. 

We also calculate $\theta_{BN}$ angles at each spacecraft and show that these can be very different at different points on its surface (Table~\ref{tab:mva}). This should be taken into account in any interpretation of data from IP shocks.

Our findings fit well with the 2D  hybrid simulation results which show that:
\begin{itemize}
\item{} Shock profiles and local $\theta_{BN}$ may vary significantly at separations of the order of 5~d$_i$ and more.
\item{} At any given time, different locations on the shock surface exhibit values of $\theta_{BN}$ anywhere between 10$^\circ$ and 90$^\circ$. 
\item{} The geometry at the particular point on the shock surface varies with time and the location on the shock surface. 
\item{} The cause of irregularities of observed shocks may be upstream magnetic structures that are convected towards the shocks. These can arise due to small amount of backstreaming ions reflected by the shocks. 
\end{itemize}

The fact that irregularities of simulated shock surfaces  may occur at quite small spatial scales is not reflected in our observations (Figure~\ref{fig:scatter}), possibly due to low number of our case studies.

In the supplement repository and Appendix B we show that fluctuations in the ULF frequency range (0.01-0.1~Hz) exist upstream of all four shocks (Figure~1), although in the case of February 2012 shock their compressive component is weak. These fluctuations could be responsible for irregular structure of observed shocks.

There is a possibility that the upstream B-field fluctuations have already been present in the upstream SW and the shocks just caught up with them. In the supplement repository and Appendix B we show ion spectra for the January 2001 and April 2010 shocks in Figure 2. The ion particle energy flux peaks at shock transitions, suggesting the ions are accelerated by the IP shocks (for the other two shocks the data were not available). In the case of the April 2010 shock part of these ions and ULF fluctuations may have actually come from the Earth's bow shock, as the data suggest that the Cluster probes have entered and exited it on several occasions before the IP shock arrival. These excursions are marked by increased suprathermal ion fluxes (green trace in the middle panel of the Figure 2b). The last excursion into the foreshock occured $\sim$20 minutes prior to the IP shock arrival and may have lasted some minutes after the shock was observed.

 \begin{figure}[h]
 \centering
 \includegraphics[width=1.\textwidth]{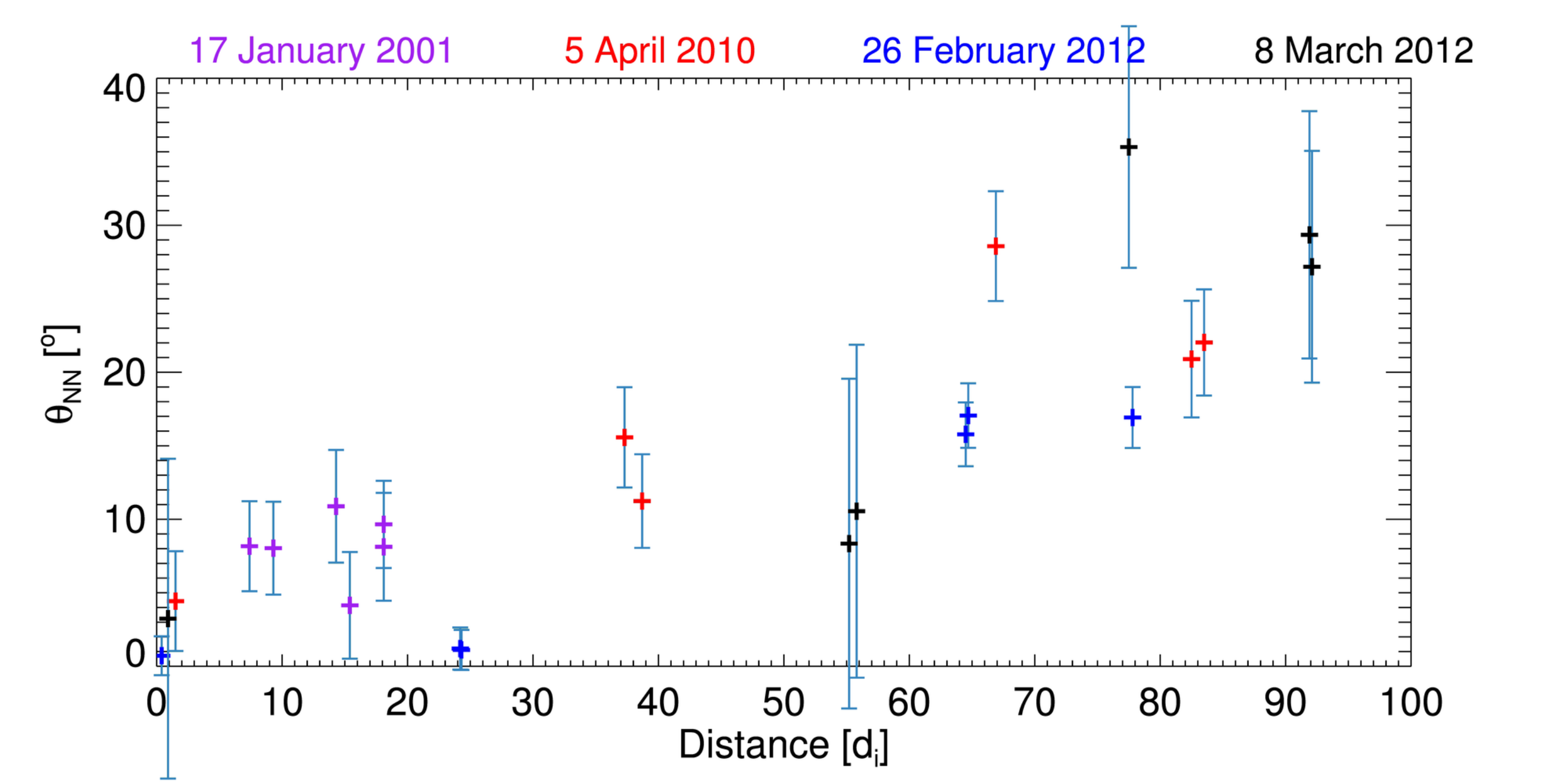}
 \caption{Angles between pairs of local shock normals shown as a function of inter-spacecraft separations in units of ion inertial lengths. The error bars show the standard deviation of $\theta_{NN}$.}
 \label{fig:scatter}
 \end{figure}
  
 \begin{figure}[h]
 \centering
 \includegraphics[width=1.0\textwidth, angle = 0]{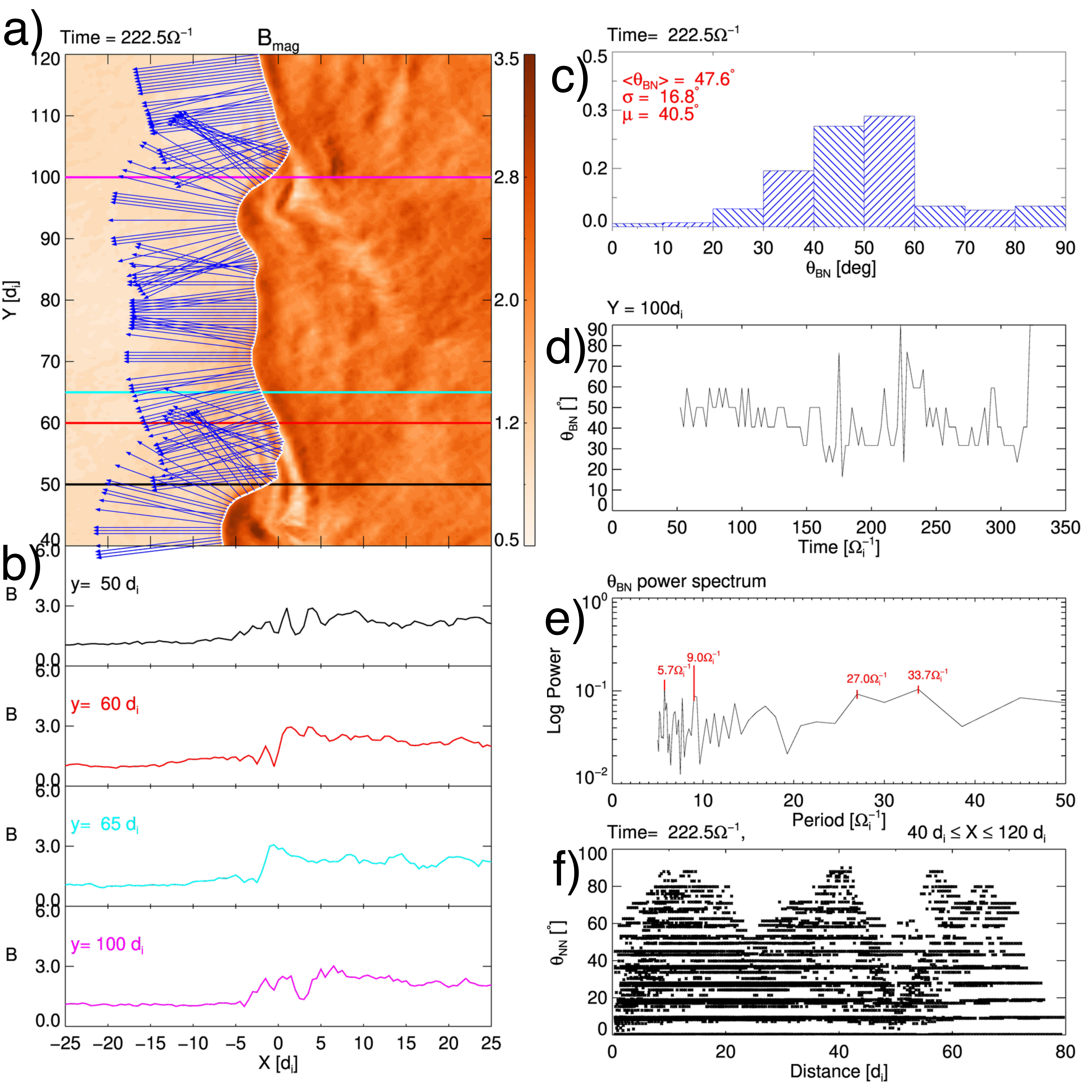}
 \caption{Simulation results from the 2D HYPSI run at the simulation time t=222.5~$\Omega_i^{-1}$.
a) B-field magnitude. $x$=0 is the coordinate of the shock obtained from the average (over $y$) B-field profile. Horizontal lines show the coordinates of the B-field profiles shown on panel b. White curve marks the shock surface.
b) B-field profiles at y=50~d$_i$ (black), 60~d$_i$ (red), 65~d$_i$ (cyan) and 100~d$_i$ (magenta).
c) Histogram of all angles $\theta_{BN}$ on the shock surface at t=11.25~$\Omega_i^{-1}$. 
d) Evolution of $\theta_{BN}$ for a point on the shock surface with y=100~d$_i$.
e) Power spectrum of $\theta_{BN}$.
f) $\theta_{NN}$ as a function of distance between pairs of normals shown on panel a). 
}
 \label{fig:simulation}
 \end{figure} 

\begin{table}
\resizebox{\textwidth}{!}{%
\centering
\begin{tabular}{c c c c c c c c c}
\hline
Spacecraft &17 Jan 2001 & & 5 Apr 2010 & & 26 Feb 2012 & & 8 Mar 2012\\
\hline
 & $\theta_{NN} [^\circ]$ & Distance (D$_i$) & $\theta_{NN} [^\circ]$ & Distance (D$_i$) & $\theta_{NN} [^\circ]$ & Distance (D$_i$) & $\theta_{NN} [^\circ]$ & Distance (D$_i$)\\
\hline
C1-C2 &  4 $\pm$4& 15.4 &  29 $\pm$4& 66.9&  17 $\pm$2& 77.8&  35 $\pm$8& 77.5\\
C1-C3 &  8 $\pm$3& 9.3 &  16 $\pm$3& 37.3&   1 $\pm$1& 24.2&  29 $\pm$8& 91.9\\
C1-C4 & 10 $\pm$4& 14.3&  11 $\pm$3& 38.7&   1 $\pm$1& 24.3&  27 $\pm$8& 92.1\\
C2-C3 & 10 $\pm$3& 18.1&  21 $\pm$4& 82.5&  16 $\pm$2& 64.5&   8 $\pm$11& 55.2\\
C2-C4 &  8 $\pm$4& 18.1&  22 $\pm$4& 83.2&  17 $\pm$2& 64.7&  11 $\pm$11& 55.8\\
C3-C4 &  8 $\pm$3& 7.4&   4 $\pm$3& 1.48&   1 $\pm$1& 0.37&   3 $\pm$11& 0.92\\
\hline
\multicolumn{8}{c}{$\theta_{BN} [^\circ]$}\\
\hline
C1 &  19& &  14& &  15& &  57& \\
C2 &  22& &  21& &  15& &  21& \\
C3 &  30& &  14& &  10& &  32& \\
C4 &  26& &  10& &  10& &  30& \\
\hline
\end{tabular}}
\caption{Results from local shock normals calculations by using the maximum variance analysis. Top: angles between pairs of normals. Bottom: $\theta_{BN}$ angle at each spacecraft.}
\label{tab:mva}
\end{table}

{\it Acknowledgments\\}
Authors acknowledge ClWeb and Cluster Science Archive teams for easy access and visualization of the data.  PK's work was supported by PAPIIT grant IA101118. XBC's work was supported by PAPIIT IN-105218 and Conacyt 255203 grants. DT acknowledges support of a studentship funded by the Perren Fund of the University of London. The authors acknowledge support from the Royal Society Newton International Exchange Scheme (Mexico) grant NI150051. Simulations were run on MIZTLI supercomputer in the frame of DGTIC LANCAD-UNAM-DGTIC-337 grant.

\section{Appendix A}
\label{appendixA}
In this section we explain how the shock normals were calculated and how the errors of $\theta_{NN}$ and $\theta_{BN}$ were estimated.

There are two main sources of errors. The first is the error of the MVA method itself which depends on the number of measurement points and the calculated eigenvalues \citep[][]{sonnerup:1998}:

\begin{equation}
\theta_{Err} = \sqrt{\frac{\lambda_3}{M-1}\frac{\lambda_2}{\lambda_2-\lambda_3}}.
\end{equation}

\begin{figure}[h]
 \centering
 \includegraphics[width=0.8\textwidth]{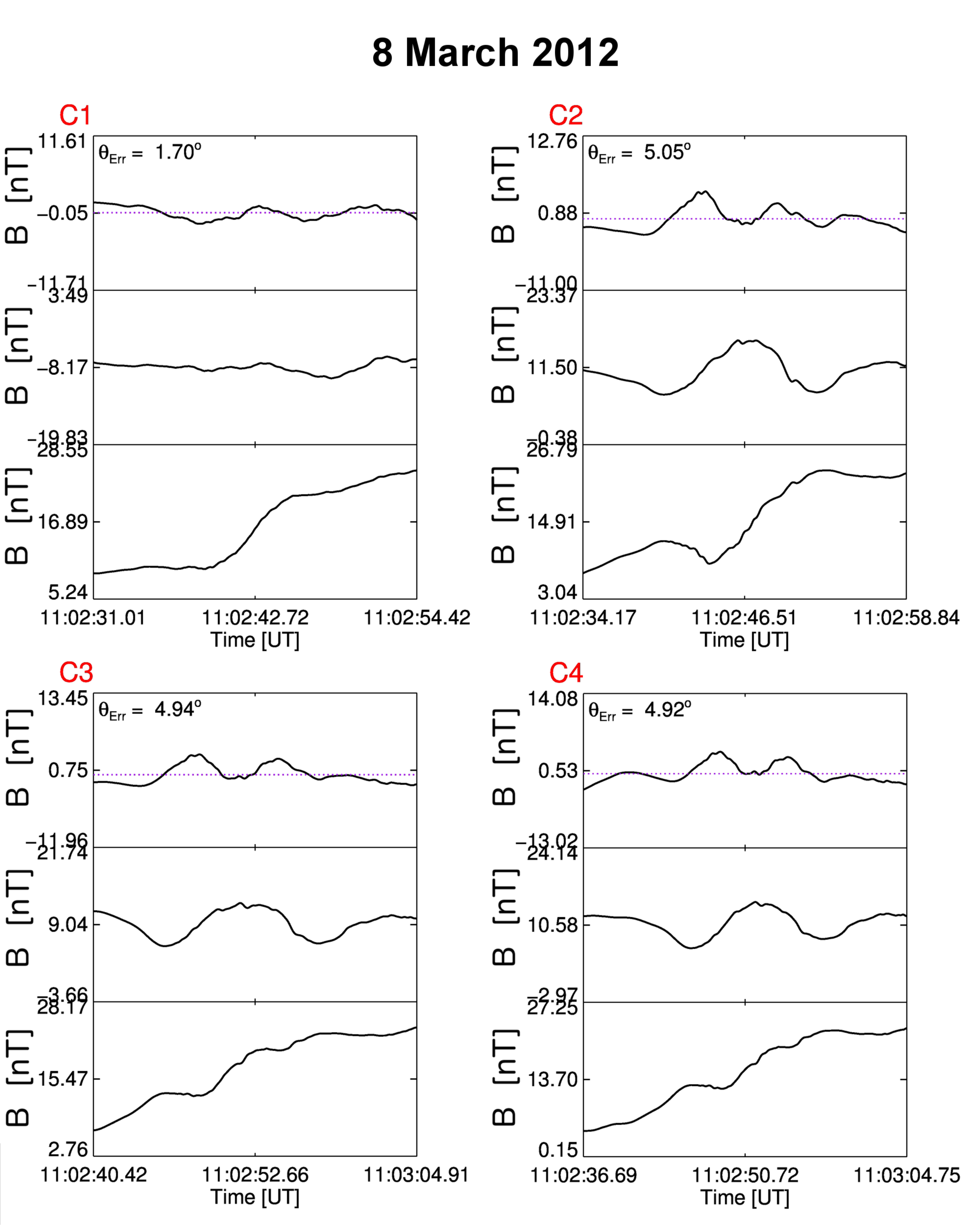}
 \caption{Magnetic field profiles of the March 2012 shock. The B-field components are in the shock-normal coordinate system. The dotted horizontal lines indicate zero value. All 160 profiles may be seen in the supplement located on-line at http://usuarios.geofisica.unam.mx/primoz/IPShockRipplingSupplement/SNCS.pdf}
 \label{fig:snc}
 \end{figure}

Here $\lambda_2$, $\lambda_3$ and $M$ are the intermediate and minimum eigenvalues and the number of measurement points, respectively. This error is stated in the Figure~\ref{fig:snc}.

The second source of errors comes from determining time intervals which are used for the MVA. These intervals need to include the shock transition but also some upstream and downstream regions. One needs to select the intervals carefully so not to include large B-field rotations that are not associated with shocks and could affect the the determination of the direction of maximum variance. We select the time intervals by hand. We repeat the process for each shock and spacecraft ten times. We then proceed to calculate angles between pairs of normals from different spacecraft ($\theta_{NN}$) and calculate the the average angles and the error of the mean. Next we sum this error with $\theta_{Err}$ in order to estimate the total error of our method. The latter is shown in Table~\ref{tab:mva} and in Figure~\ref{fig:scatter} in form of error bars.

The $\theta_{BN}$ errors stem from the MVA method and the selection of the upstream time intervals over which we calculate the average B-field direction. They are typically $\sim$15~seconds long. After repeating this selection for 10 times we estimate the errors to be $\sim$5$^\circ$.

\section{Appendix B}
\label{appendixB}
Here we present a) wavelet spectra of magnetic field fluctuations observed by Cluster~1 spacecraft upstream of the four interplanetary (IP) shocks (Section~\ref{wavelet}); b) ion spectrograms and energy fluxes around two of the shocks for which the data were available (Section~\ref{particles}); and c) Simulation results of our high-M$_A$, low-$\beta$ run (Section~\ref{simulation}).

\subsection{Wavelet spectra of upstream waves}
\label{wavelet}

\begin{figure}[h!]
\includegraphics[width = 1.0\textwidth]{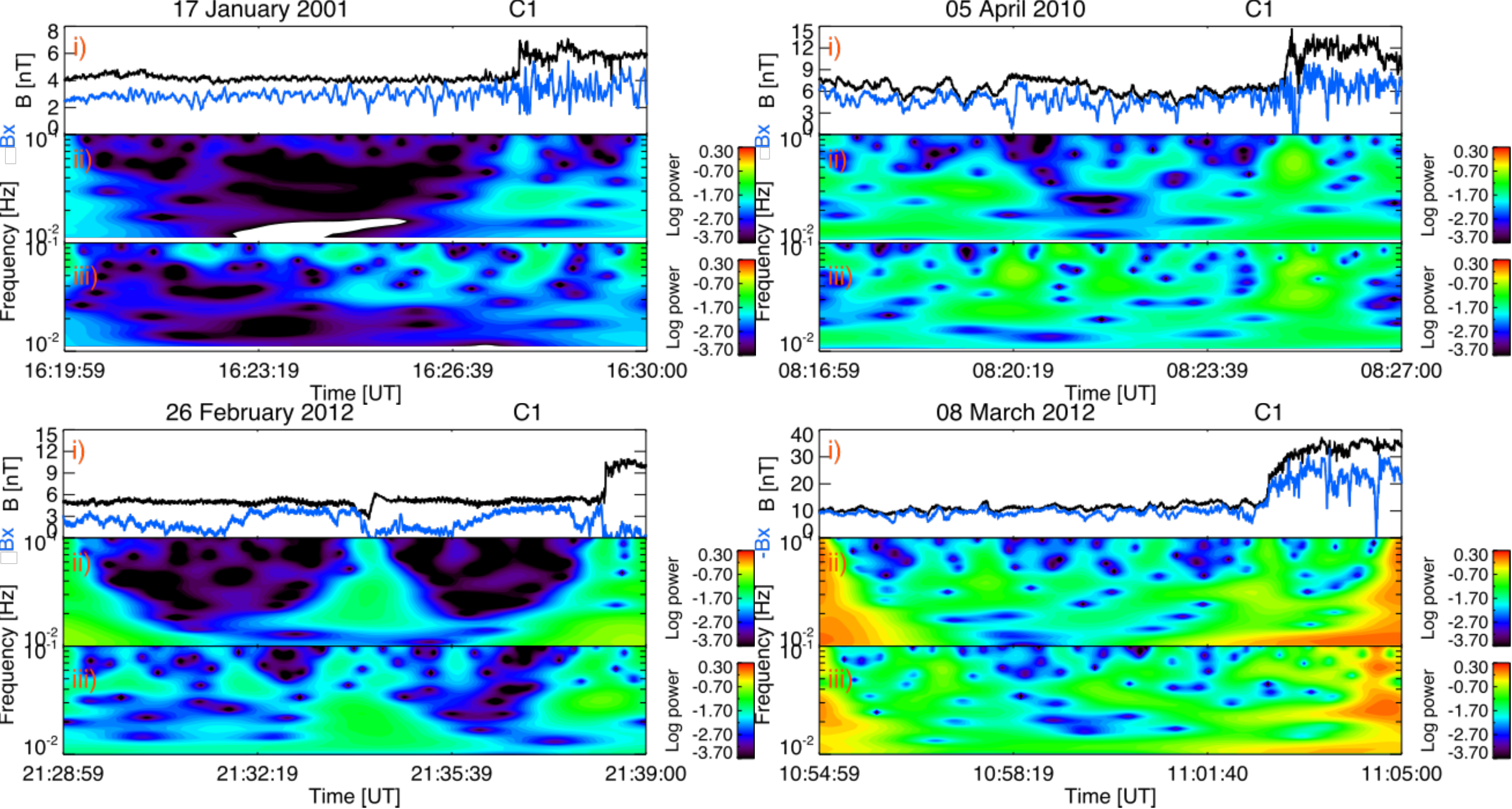}
\caption{Magnetic field data and wavelet spectra during time periods when the four IP shocks were observed. Black traces on panels i) represent the magnetic field magnitude. Blue traces on panels i) show B$_{x,GSE}$ or -B$_{x,GSE}$ magnetic field component. Panels ii) and iii) exhibit wavelet spectra of the B and B$_x$, respectively.}
\label{fig:wavelet}
\end{figure}

Figure~\ref{fig:wavelet} shows magnetic field data and the corresponding wavelet spectra for the four IP shocks observed on 17 Januar 2001, 5 April 2010, 26 February 2012 and 3 March 2012. Panels i) show B-field magnitude data as black lines, while B$_{x,GSE}$ or -B$_{x,GSE}$ component is represented by the blue line. Panels ii) and iii) exhibit B and B$_{x,GSE}$ wavelet spectra, respectively. We can see that compressive and/or transverse B-field fluctuation in the frequency range 10$^{-2	}$-10$^{-1}$~Hz are present upstream of all four shocks. In general, there is more power in the transverse component of these fluctuations than in the compressive component.	 

\subsection{Particle data}
\label{particles}

\begin{figure}[h!]
\includegraphics[width = 0.7\textwidth]{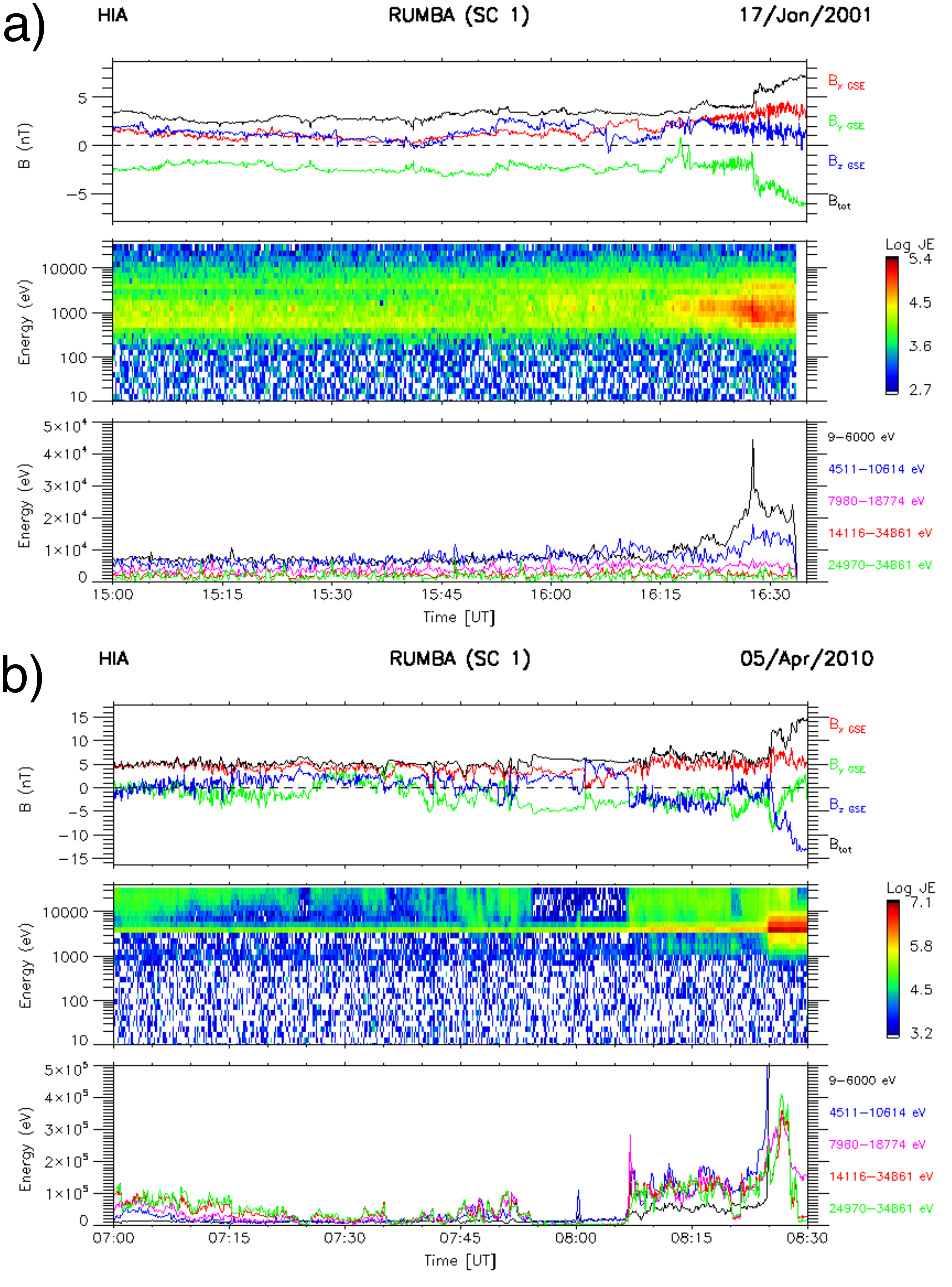}
\caption{Panels a) and b) show magnetic field data from FGM and particle spectrograms and fluxes for 17 January 2001 and 5 April 2010 shocks.}
\label{fig:particles}
\end{figure}

Figure~\ref{fig:particles} shows magnetic field, particle spectrogram and particle energy fluxes at time when a) 17 January 2001 and b) 5 April 2010 shocks were observed. In both cases the suprathermal ion energy fluxes in units (in units of keV/(s$\cdot$cm$^{-2}\cdot$sr$\cdot$keV)) start increasing before the shock arrival and peak at shock transition, suggesting they are accelerated by the shocks. The suprathermal ion energy flux (and magnetic ULF fluctuations) in the case of the 5 April 2010 could partially arrive from the Earth's bow-shock, since the ion spectrogram suggests that prior to and possibly during the shock encounter, the Earh's foreshock has been observed intermittently.

\subsection{Simulation results for high-M$_A$, low-$\beta$ run.}
\label{simulation}

\begin{figure}[h!]
\includegraphics[width = 1.0\textwidth]{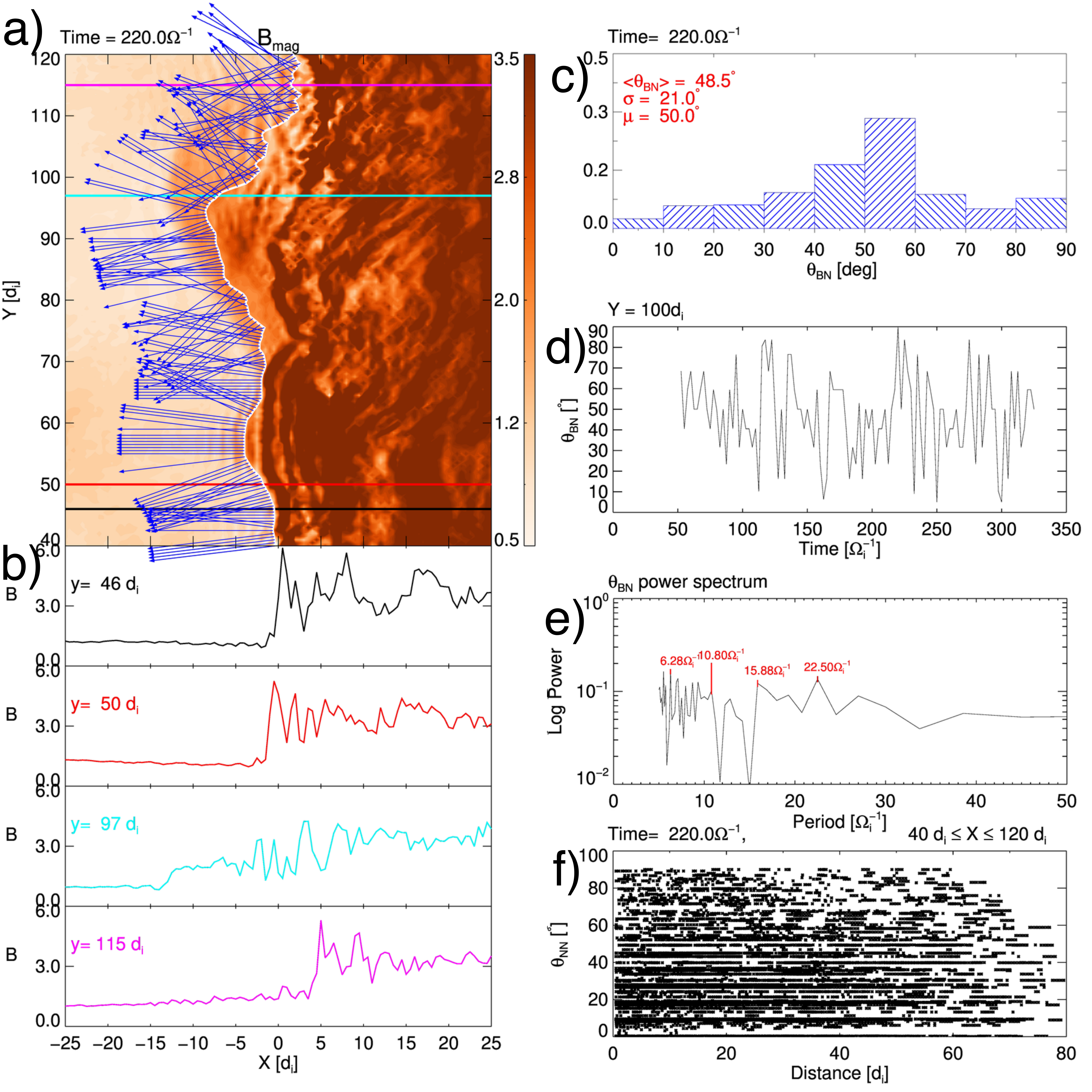}
\caption{Results from our high-M$_A$, low-$\beta$ run.}
\label{fig:simulation2}
\end{figure}

Figure~\ref{fig:simulation2}a shows results from our high-M$_A$ (=6.5), low-$\beta$ (=0.2) run at time t=112.5~$\Omega_i^{-1}$) with x = [-25,25]~d$_i$ and y = [40, 120]~d$_i$. $x$=0 is the average shock position obtained from the averaged (in $y$ direction) B-field profile. The colors represent the B-field magnitude. The white curve marks the shock front and blue arrows show the directions of local shock normals. 

Figure~\ref{fig:simulation2}b shows four B-field profiles for $y$=46~d$_i$ (black), 50~d$_i$ (red), 97~d$_i$ (cyan) and 115~d$_i$ (magenta). Animations for this run can be found at \\
http://usuarios.geofisica.unam.mx/primoz/IPShockRipplingSupplement/ and are titled BfieldLowBeta.avi, BfieldHighLowShort.avi. 

Figure~\ref{fig:simulation2}c shows the distribution of $\theta_{BN}$ angles at time t=112.5~$\Omega_i$. 

Figure~\ref{fig:simulation2}d shows the time evolution of the $\theta_{BN}$ for the point on the shock surface at y=97~d$_i$. 

Figure~\ref{fig:simulation2}e shows $\theta_{NN}$ angles for pairs of normals shown on panel a). 

\end{document}